\documentstyle[aps,twocolumn,prl,epsfig]{revtex}
\begin{document}

\wideabs{
  
\author{Juan J. Garc\'{\i}a-Ripoll$^1$, V\'{\i}ctor M.
  P\'erez-Garc\'{\i}a$^1$, Wieslaw Krolikowski$^2$, and Yuri S. Kivshar$^3$}

\address{$^1$Departamento de Matem\'aticas, E. T. S. I. Industriales, Universidad de Castilla-La Mancha, 13071 Ciudad
  Real, Spain \\
  $^2$Laser Physics Centre, The Australian National University, Canberra,
ACT 0200,
  Australia \\
  $^3$Optical Sciences Centre, The Australian National University,
Canberra, ACT 0200,
  Australia}

\title{Scattering of light by molecules of light}

\maketitle

\draft
\begin{abstract}
  We study the scattering properties of optical {\em dipole-mode vector
  solitons} recently predicted theoretically and generated in a laboratory.
  We demonstrate that such a {\em radially asymmetric} composite self-trapped
  state resembles {\em ``a molecule of light''} which is extremely robust,
survives a wide range of collisions, and displays new phenomena such as the
  transformation of a linear momentum into an angular momentum, etc.  We
  present also experimental verifications of some of our predictions.
\end{abstract}

}

Since the begining of the history, physics has studied simple objects and the
way they arrange to form more complex objects. Some remarkable success included
the atomic theory of matter, the discovery of the structure of nucleus in terms
of protons and neutrons and, more recently, the substructure of nucleons in
terms of quarks.  Such concepts seem to be restricted to ``solid" objects only,
and at a first sight it could seem that nothing similar is possible for light.
However, elementary {\em robust} objects made of light have been known since
the 70's. In fact, {\em spatial optical solitons}--- self-trapped states of
light with particle-like properties--- have attracted a considerable attention
during last years as possible building blocks of all-optical switching devices
where light is used to guide and manipulate light itself \cite{steg}.  Recent
progress in generating spatial optical solitons in various nonlinear bulk media
allows to study truly two-dimensional self-trapping of light and different
types of interactions of multi-dimensional solitary waves \cite{science}.

Robust nature of spatial optical solitons as self-trapped states of light that
they display in interactions \cite{science}, allows to draw an analogy with
atomic physics treating spatial solitons as ``atoms of light''.  Furthermore,
when several light beams generated by a coherent source are combined to produce
{\em a vector soliton}, this process can be viewed as the formation of
composite states or ``molecules of light''.

Recently, we have predicted theoretically the existence of a robust ``molecule
of light'', {\em a dipole-mode vector soliton} (or ``dipole'', for simplicity)
that originates from trapping of a dipole beam by an effective waveguide
created by a mutually incoherent fundamental beam \cite{us}.  The first
observation of this novel type of optical vector soliton has been recently
reported in Ref. \cite{exper}, where the dipoles have been generated using two
different methods: a phase imprinting and a symmetry-breaking instability of a
vortex-mode composite soliton \cite{moti}.

The concept of vector solitons as ``molecules of light'' should be compared
with photonic microcavity structures, micrometer-sized ``photonic quantum
dots'' that confine photons in such a way that they act like electrons in an
atom \cite{photonic}.  When two of these ``photonic atoms'' are linked
together, they produce a ``photonic molecule'' whose optical modes bear a
strong resemblance to the electronic states in a diatomic molecule like
hydrogen \cite{photonic2}. Self-trapped states of light we study here can be
viewed as the similar photonic structures where, however, the photonic trap and
the beam it guides are both made of light creating {\em self-trapped photonic
  atoms and molecules}.

In this Letter we study the scattering properties of dipole-mode vector
solitons (``molecules of light") and analyze, in particular, the interaction
between these objects and other robust structures made of light: scalar
solitons (``atoms of light'') and other dipoles. We describe a number of
interesting effects observed in such interactions, e.g. the absorption of a
soliton beam by a dipole and replacement of the soliton with a dipole
component, transformation of a linear momentum into an angular momentum with
subsequent dipole spiraling, etc. Additionally, we verify experimentally some
of these predictions for composite spatial solitons generated in a
photorefractive crystal. The versatility of phenomena described here makes the
dipole-mode vector soliton a complex object with promising applications in
integrated optics in addition to its fundamental interest.

{\em The model and dipole solitons.}  We consider the propagation of two
coherent light beams interacting incoherently in a saturable nonlinear medium.
In the paraxial approximation, the beam mutual interaction can be described by
a system of two coupled nonlinear Schr\"odinger (NLS) equations
\cite{us,exper,moti,spiral},
\begin{mathletters}
\label{Model}
\begin{eqnarray}
i \frac{\partial  u}{\partial z} = -\frac{1}{2} \triangle_{\perp} u + F(I) u, \\
i \frac{\partial  v}{\partial z} = -\frac{1}{2} \triangle_{\perp} v + F(I) v,
\end{eqnarray}
\end{mathletters}
where $u({\bf r}_{\perp}, z)$ and $v({\bf r}_{\perp}, z)$ are dimensionless
envelopes of the beams self-trapped in the cross-section plane ${\bf r}_{\perp}
= (x,y)$ and propagating along $z$.  The function $F(I) = I (1 +s I)^{-1}$
characterizes a saturable nonlinearity of the medium, where $s$ is a
dimensionless saturation parameter $(0 < s <1)$ and $I = |u|^2 + |v|^2$ is the
total intensity.

Equations (\ref{Model}) describe different types of spatially localized
composite solutions.  {\em The dipole-mode vector soliton} (or ``a molecule of
light'') is a stationary state which is composed of a nodeless beam in the $v$
component and a dipole beam (or a pair of out-of-phase solitons) in the $u$
component.  Solitons in the $u$ component have opposite phases and thus they
tend to repel each other, but the role of the beam $v$ is to stabilize the
structure making it robust.  A numerical analysis of the linearized equations
(\ref{Model}) shows no signs of linear instability of this composite structure
\cite{us}.  Moreover, it was shown \cite{us} that such robust dipole-mode
vector solitons exist for a wide range of the beam powers $P_u=\int |u|^2 d
{\bf r}_{\perp}$ and $P_v=\int |v|^2 {\bf r}_{\perp}$.  Since we are interested
in showing stability {\em far from the regime in which one beam is dominant and
  acts as a waveguide}, all numerical experiments are performed using as
initial states for the collisions experiments stationary states in which $P_u
\simeq P_v$.

%
%

{\em Qualitative analysis.}  We are interested in the dynamics of the dipole
under the action of finite external perturbations introduced by its collision
with other objects.  The word ``finite'' emphasizes the fact that we can no
longer make use of linearized equations and that we must deal with the full
system (\ref{Model}). This fact, combined with the complex structure of the
dipole which lacks radial symmetry, makes analytical predictions on the dipole
dynamics very difficult. Nevertheless, as will be shown below, one may extract
some general rules on which qualitative predictions may be based.

The idea is that the dipole can be seen as a bound state of a soliton beam
(in $u$) plus a pair of vortices with  opposite charges (in $v$), and
therefore many effects observed in the beam collisions can be understood
once the mutual interaction of these simpler objects is known.

One of the components of the dipole is a soliton beam (to be referred to as
{\em soliton} hereafter). Spatial solitons are stable localized states
which have no nodes and which are the states of minimum energy of the
system {\em for a fixed power}.  When two of these solitons are in
different beams (say,  one in $u$ and the other one,  in $v$), they interact
incoherently and {\em attract each other}. Thus, during an incoherent
interaction two solitons {\em attract} each other and either become bound
or scatter. In the former case, we have an example of what we call a {\em
molecule of light}, which is typically referred to as ``vector soliton''.
However, when two solitons are in the same component, their mutual
interaction depends on their phase difference.  When this quantity is small
or zero,  they {\em interact attractively}, whereas if their mutual phases
differ by $\pi$, they {\em repel each other}.

Another nonlinear structure that should be mentioned in this context is
{\em a vortex-mode composite soliton} introduced in \cite{moti}
which in our model (\ref{Model})  is an unstable object (see details 
in \cite{us}).  Vortices may only
be partially stabilized (i.e., their decay rate becomes smaller) by sharing
space with a large soliton beam (e.g., when a vortex  in the beam $u$ is
guided by an effective waveguide created in the component $v$).  Thus, a
composite state of a vortex plus soliton constitutes an unstable  molecule
of light.

Concerning a dipole, it can be seen as a pair of vortices as described
above or,  alternatively, as  a bound state of two solitons with a phase
difference of $\pi$. Although, in principle,  these solitons should repel
each other, the system is stabilized due to the interaction with a
soliton-induced waveguide created  in the other component.

%
%

{\em Soliton-dipole scattering.} The first type of numerical simulations we
present here consists in shooting a scalar soliton against a dipole-mode vector
soliton.  All the simulations discussed here have been performed using a
split-step operator technique using FFT, with grid sizes of up to
512$\times$512 points covering a rectangular domain of 68$\times$34
adimensional units.  The initial data are always a combination of stationary
states. For instance, when a soliton is launched against a dipole, we start
with
\begin{mathletters}
\begin{eqnarray}
u({\bf x},0) & = & u_{dipole}({\bf x}) + u_{soliton}({\bf x}-{\bf d})
e^{-i{\bf p_0 x}},\\
v({\bf x},0) & = & v_{dipole}({\bf x}).
\end{eqnarray}
\end{mathletters}
Here ${\bf d}=(d_x,d_y)$, $d_x \gg d_y$, $d_y$ is {\em the impact
parameter},  and
$\mathbf{p_0}$ is proportional to the initial (linear) momentum of the
incoming scalar soliton.
The initial data $u_{dipole}$, $u_{soliton}$,  and $v_{dipole}$ are
obtained numerically by a suitable minimization procedure outlined in Ref.
\cite{us}.

%
%

\begin{figure}
\begin{center}
\epsfig{file=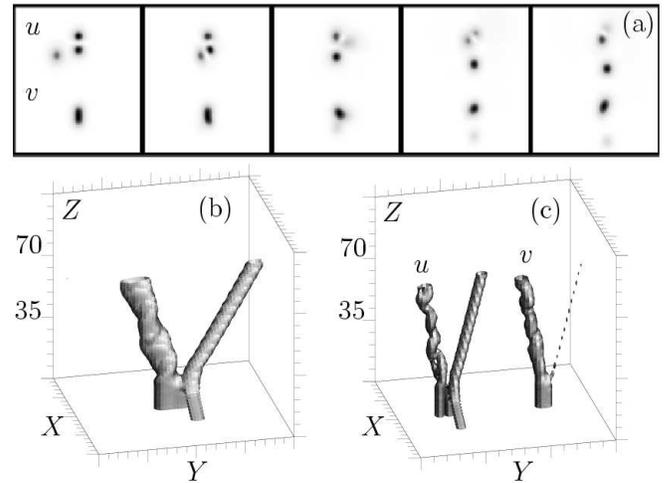,width=\linewidth}
\end{center}
\caption{
  \label{fig-scat}
  Soliton-dipole scattering. (a) Snapshots of the intensity profile of each
  beam.  (b) 3D plot of the total intensity $|u|^2+|v|^2$, which shows the
  rotation induced in the dipole. (c) Same as (b), but with separated $u$ and
  $v$ components.}
\end{figure}

The result is an inelastic collision in which the soliton becomes deflected and
the dipole gains both {\em linear} and {\em angular} momenta. The whole process
is depicted in Fig. \ref{fig-scat}. Soliton scattering occurs when the incident
beam has medium to large linear momentum or when it has an appropriate initial
phase. For instance, in Fig. \ref{fig-scat} the incident soliton has sign $(-)$
and it crashes against the part of the dipole with $(+)$ sign. A conservation
law forces the dipole to rotate and the soliton becomes deflected, sometimes as
much as by a 90 degrees angle.

\begin{figure}
\begin{center}
\epsfig{file=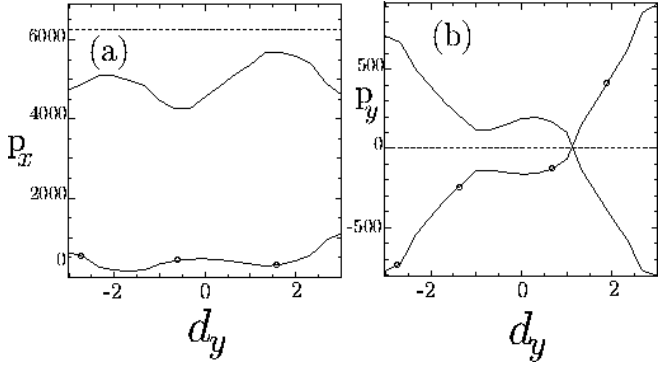,width=\linewidth}
\end{center}
\caption{
  \label{fig-P}
  (a, b) Components of the linear momentum of the incident soliton (solid line)
  and dipole (marked by circles) after an inelastic collision with a large
  incident momentum, $\mathbf{p_u} \equiv \int u^* \nabla u \, d{\bf
    r}_{\perp}$, as a function of the impact parameter $d_y$, which shows the
  crucial role of the dipole asymmetry. Total $p_y$ is not zero because of
  resultant radiation which is not seen in the figure.}
\end{figure}

\begin{figure}
\begin{center}
\epsfig{file=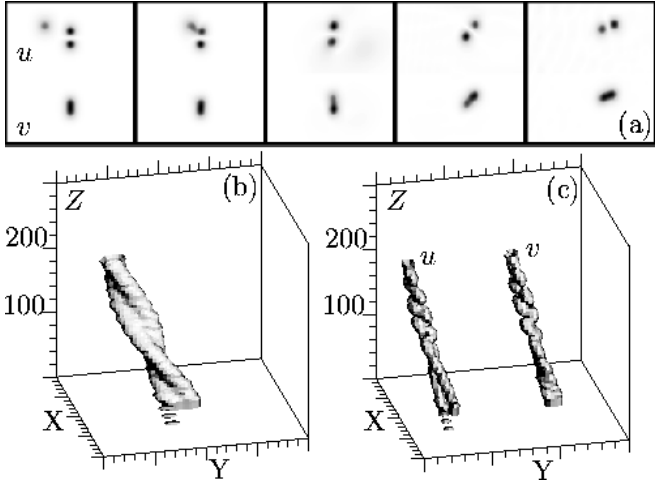,width=\linewidth}
\end{center}
\caption{
  \label{fig-abs}
  Absorption of a soliton by a dipole: (a) an intensity profile of
  each light beam --the darker the more intense; (b) 3D plot of
  the total beam intensity; (c) 3D plots where both $u$ and $v$ components
have been separated.}
\end{figure}

When the linear momentum of the incident soliton is large, it moves too fast to
suffer a destructive influence from the dipole. In Fig.  \ref{fig-P} we plot
the exchange of the linear momentum between the soliton and dipole as a
function of the impact parameter. The effective interaction is clearly {\em
  attractive}: the soliton coming from below ($d_y < 0$) feels the drag of the
dipole above it and gets deflected upwards ($p_y > 0$), while the dipole moves
downwards.

%
%

{\em Soliton absortion by a dipole.} The second family of numerical experiments
is performed with solitons which are slow and, as is usual in scattering
processes, the effects of the interaction process may be more drastical. For
some impact parameters the soliton gets too close to the lobe of the dipole
with the smallest phase difference and fuses with it with some emission of
radiation and a subsequent rotation of the dipole.  This is well reflected in
Fig.  \ref{fig-abs} (radiation is not seen).

%
%

\begin{figure}
\begin{center}
\epsfig{file=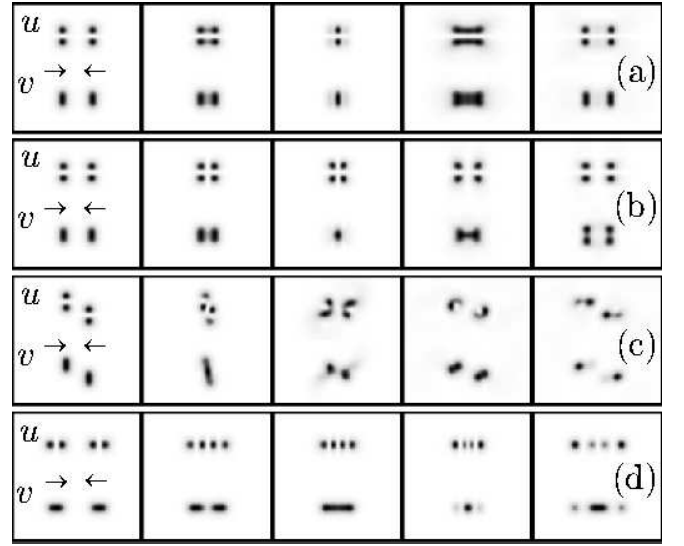,width=\linewidth}
\end{center}
\caption{
  \label{fig-dd}
  Collisions of two dipoles against each other, seen from the center of mass
  of the system.
  }
\end{figure}

{\em Dipole-dipole collisions.} The third family of numerical simulations
corresponds to shooting dipoles against each other. These collisions provide a
rich source of phenomena depending on the mutual orientation of the dipoles and
on the initial energy. Figure \ref{fig-dd} summarizes the main results
observed.  There we see {\em three cases} (a-c) in which the dipole solitons
are preserved. The figure shows an in-phase collision with weak interaction
[Fig. \ref{fig-dd}(a)], an out-of-phase collision with repulsion [Fig.
\ref{fig-dd}(b)], and an example of the collision with nonzero impact parameter
in which two vortex states are created and they decay into a pair of spiralling
solitons [Fig. \ref{fig-dd}(c)].
The last case, Fig. \ref{fig-dd}(d), shows an interesting inelastic process
when two dipoles fuse into a more complex state which then decays creating a
new dipole and a pair of simple solitons.  All these processes may be
understood in terms of the phase of the lobes of each dipole as described
above.

%
%

\begin{figure}
\begin{center}
\epsfig{file=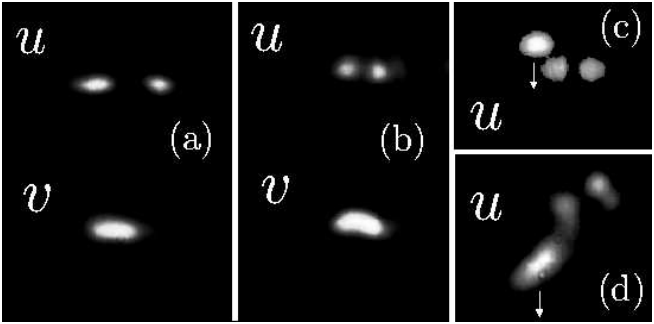,width=\linewidth}
\end{center}
\caption{Experimental demonstration of the soliton-dipole collisions. (a,b)
  Formation of a dipole-mode soliton from two out-of-phase coherent beams ($u$)
  and co-propagating beam ($v$): (a) input beams, (b) output beams.
  Experimental parameters are: $V = 1.8$ kV, $z =10$ mm, and the initial powers
  are $P_v = 2 \, \mu$W, $P_u = 2.6 \, \mu$W. (c,d) Soliton-dipole interaction
  shown prior (c) and after (d) the collision, only the dipole component of the
  composite soliton is shown. Rotation of the dipole-mode vector soliton after
  the interaction is clearly visible.
  \label{fig-exp}
  }
\end{figure}

{\em Experimental results.} Generation of an isolated dipole-mode vector
soliton was reported earlier in Ref. \cite{exper}.  The dipole-mode soliton was
created using two different processes: (i) {\em phase imprinting}, when one of
the beam components is sent through a phase mask in order to imprint the
required phase structure, and (ii) {\em symmetry-breaking instability} of a
vortex-mode composite soliton.  In this way, we obtain a dipole-like structure
with a phase jump along its transverse direction that is perpendicular to the
optical axis of the crystal [see Fig. \ref{fig-exp}(a), the beam $u$].  That
dipole-like beam is then combined with the second, nodeless beam [the beam $v$
in Fig.  \ref{fig-exp}(a)], and the resulting composite beam is focused into
the input face of the photorefractive SBN crystal (the crystal has the same
parameters as in \cite{exper}), biased with the DC field of 1.5-2.5 kV applied
along an optical axis.  To control the degree of saturation, we illuminate the
crystal with a wide beam derived from a while light source. Propagating in an
effectively {\em self-focusing saturable medium}, such a composite input beam
creates a dipole-mode vector soliton, as shown in Fig. \ref{fig-exp}(b) (both
components are shown separately).  As discused above, the beam $v$ creates an
effective asymmetric waveguide that guides a dipole-like mode in the form of
two out-of-phase solitary beams that mutually repell each other.

To observe the soliton-dipole interaction effects, we launch a scalar soliton
beam against the dipole soliton. The input state is shown in Fig.
\ref{fig-exp}(c), where the dipole-mode soliton is presented by its two-lobe
$u$ component only.  When the soliton interacts with a dipole, it gets
deflected and transforms a part of its linear momentum into an angular momentum
of the dipole that starts rotating is visible in Fig. \ref{fig-exp}(d). A
qualitative comparison between the theory and experiment is hard to carry out
since the original theoretical model is isotropic, while the bias
photorefractive crystal is known to possess an anisotropic nonlocal
nonlinearity \cite{wieslaw} making the dipole rotation {\em anisotropic}, since
the dipole structure along and perpendicular to the optical axis is different.

{\em Conclusions.} We have studied the phenomenology of collisions of the
recently discovered dipole-mode vector solitons with other nonlinear localized
structures.  Apart from checking the robustness of the dipoles against strong
interactions, we have shown that in many cases they behave qualitatively as
tightly bound molecules of light, with {\em two major degrees of freedom}
(rotation as a whole and oscillation of the lobes of the dipole) which can be
excited by collisions.  Sometimes the dipole excitation is so strong that the
structure behaves as a pair of spiraling beams earlier analyzed in Ref.
\cite{spiral}.  This is only one of many interesting phenomena observed in
simulations which also include excitation of rotational modes by collision with
a scalar soliton, annihilation or strong deflection of the incident soliton,
etc. Even richer phenomenology is observed when two dipole-mode solitons are
made to collide.  It is remarkable that the rich behavior observed here may be
understood qualitatively in terms of the structure of the objects colliding and
the relative phases of the dipole components. Finally, we have also verified
experimentally some of our predictions.

A rich variety of the effects described here might make these objects good
candidates for practical applicability in the field of integrated optics. In
this sense, the dipole-mode vector soliton resembles the operation of an
electronic transistor since it is an asymmetric object whose response is
nonlinear and depends on the directionality of the input. We think the behavior
described here, in addition to its fundamental interest, may have wide
applicability in the future.

The work has been partially supported by a grant of the Planning and
Performance Fund grant, by APCRC, and by DGICYT (grant PB96-0534).

\end{document}